\documentclass[
    ,final            
  sort]
  {aipproc}

\layoutstyle{8x11double}

\usepackage[ansinew]{inputenc}
\usepackage[percent]{overpic}

\begin{document}



\newcommand{\actaa}{Acta Astronomica}
\newcommand{\ao}{Applied Optics}
\newcommand{\aap}{Astronomy and Astrophysics}
\newcommand{\aapr}{Astronomy and Astrophysics Reviews}
\newcommand{\aj}{The Astronomical Journal}
\newcommand{\apj}{The Astrophysical Journal}
\newcommand{\apjl}{The Astrophysical Journal, Letters to the Editor}
\newcommand{\apjs}{The Astrophysical Journal, Supplement Series}
\newcommand{\aplett}{Astrophysics Letters and Communications}
\newcommand{\apspr}{Astrophysics Space Physics Research}
\newcommand{\apss}{Astrophysics and Space Science}
\newcommand{\aaps}{Astrophysics and Space Science, Supplement Series}
\newcommand{\araa}{Annual Review of Astronomy and Astrophysics}
\newcommand{\gca}{Geochimica Cosmochimica Acta}
\newcommand{\grl}{Geophysical Research Letters}
\newcommand{\icarus}{Icarus}
\newcommand{\mnras}{The Monthly Notices of the Royal Astronomical Society}
\newcommand{\jcp}{Journal of Chemical Physics}
\newcommand{\jcis}{Journal of Colloid and Interface Science}
\newcommand{\jfm}{Journal of Fluid Mechanics}
\newcommand{\jgr}{Journal of Geophysical Research}
\newcommand{\planss}{Planetary and Space Science}
\newcommand{\nat}{Nature}
\newcommand{\prl}{Physical Review Letters}
\newcommand{\pra}{Physical Review A}
\newcommand{\prb}{Physical Review B}
\newcommand{\prc}{Physical Review C}
\newcommand{\prd}{Physical Review D}
\newcommand{\pre}{Physical Review E}
\newcommand{\rsi}{Review of Scientific Instruments}
\newcommand{\solphys}{Solar Physics}
\newcommand{\ssr}{Space Science Reviews}
\newcommand{\zap}{Zeitschrigt f{\"u}r Astronomie}

\title[Ice-Particle Collisions in Saturn’s Rings]{Laboratory Studies of Ice-Particle Collisions in Saturn’s Dense Rings}


\author{Daniel Heißelmann}{
  address={Institut für Geophysik und extraterrestrische Physik, Technische Universität Braunschweig, Mendelssohnstraße~3, 38106 Braunschweig, Germany},
  email={d.heisselmann@tu-bs.de}
}

\author{Jürgen Blum}{
  address={Institut für Geophysik und extraterrestrische Physik, Technische Universität Braunschweig, Mendelssohnstraße~3, 38106 Braunschweig, Germany},
  email={j.blum@tu-bs.de}
}

\author{Kristin Wolling}{
  address={Institut für Geophysik und extraterrestrische Physik, Technische Universität Braunschweig, Mendelssohnstraße~3, 38106 Braunschweig, Germany},
  email={k.wolling@tu-bs.de}
}

\begin{abstract}
In this work, we report on microgravity studies of particle ensembles simulating ice-particle collisions in Saturn's dense main rings. We have developed an experimental method to study the energy dissipation in a many-body system consisting of approx. one hundred cm-sized glass spheres. The temporal development of the mean particle velocity, ranging from $\sim 10\,\mathrm{cm\,s^{-1}}$ (at the beginning) to $\sim 0.35\,\mathrm{cm\,s^{-1}}$ (after 9~s of experiment duration), can be explained by a constant coefficient of restitution $\varepsilon = 0.64$. A comparison to values obtained for pure water-ice bodies shows that future cryogenic ice-collision experiments can achieve collision velocities of $\sim 0.1\,\mathrm{cm\,s^{-1}}$, and thus will very well simulate the conditions in Saturn's main rings.
\end{abstract}

\maketitle


\section{Introduction}
Planetary rings are among the most fascinating objects in our solar system. Especially Saturn's bright rings have impressed astronomers, like G. Galilei and J.~D. Cassini, for the past four centuries. Today we know that they radially extend to several hundred thousand kilometers distance from Saturn, although they are locally confined to a thickness of only a few meters \citep{porco_et_al2008AJ, tiscareno_et_al2007Icarus, hedman_et_al2007AJ}. The radio occultation data \citep{tyler_et_al1981Science, marouf_et_al1983Icarus} obtained by deep-space missions (e.g. \emph{Voyager} and \emph{Cassini}) show that the rings can be described as `granular gases' consisting of myriads of cm- to m-sized, almost pure water-ice bodies \citep{zebker_et_al1985Icarus, poulet_et_al2003A&A} moving on Keplerian orbits. Their orbital motion is disturbed by interaction with nearby moons and so-called `moonlets' leading to an increase of orbital eccentricity, which is counteracted by the dissipation of kinetic energy in frequent inelastic collisions at relative velocities well below $1~\mathrm{cm~s^{-1}}$ \citep{esposito2002RPPh}.\par

In the past, several kinetic theories \citep{trulsen1971Ap&SS, trulsen1972Ap&SS, goldreich_tremaine1978Icarus, hameen-anttila1978Ap&SS} were developed and numerical studies \citep{schmidt_et_al2001Icarus, salo_et_al2001Icarus} were carried out to explain the dynamic phenomena, like wakes propagating throughout the rings, instabilities, and overstabilities, observed by the deep-space missions \emph{Voyager} and \emph{Cassini} \citep{hedman_et_al2007AJ, colwell_et_al2007Icarus, thomson_et_al2007GRL}. In addition, visco-elastic collision models were developed to treat fragmentation processes and mass gain of the constituent particles \citep{spahn_et_al1995CSF, brilliantov_et_al1996PRE, albers_spahn2006Icarus}.\par

In most of the simulations and kinetic theories, the energy dissipation in individual collisions is described by only one parameter, the coefficient of restitution $\varepsilon$, given by the ratio of relative velocities $v'$ (after) and $v$ (before the impact).

\section{Drop tower experiments}\label{s_exp_dt}

\begin{figure}[!tb]
  \includegraphics[width=\columnwidth]{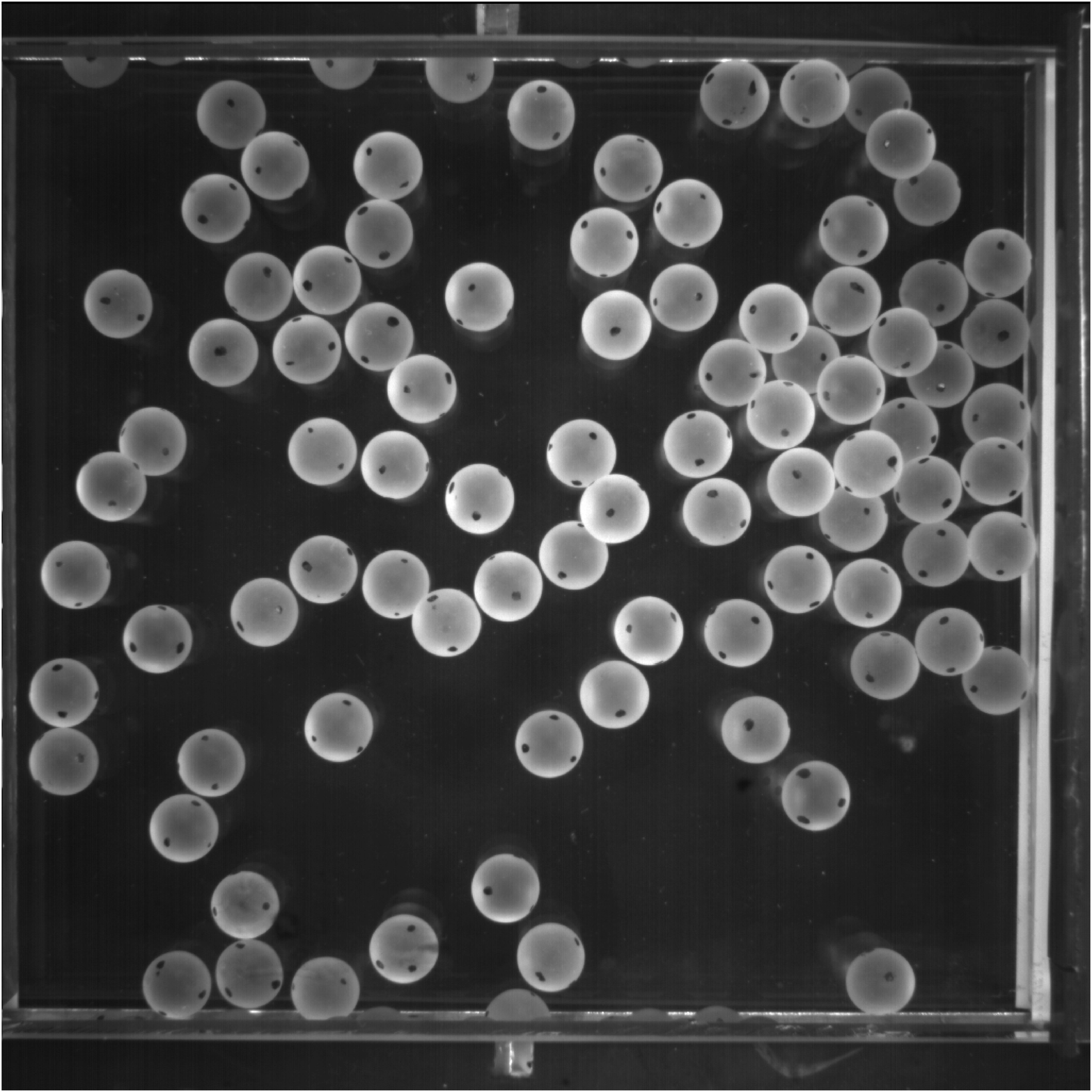}
  \caption{Individual image of a high-speed image sequence of an ensemble of 92 glass spheres (1~cm diameter) colliding in microgravity. The black marks visualize the particles' rotation that shall be treated in future analysis.}\label{f_dt_exp}
\end{figure}

We designed an experimental setup to investigate the collision processes within an ensemble of up to one hundred centimeter-sized spheres in microgravity. As a prototype, we built a glass-made test chamber of $150\times150\times15~\mathrm{mm^3}$ volume in which the sample particles can be injected from opposite sides by two electrically slid glass bars afterwards sealing the entry holes of the glass box. The typical injection velocity is $\sim 10~\mathrm{cm~s^{-1}}$. The experiments were performed at the Bremen drop-tower facility ZARM\footnote{Center of Applied Space Technology and Microgravity}, where a catapult was utilized to achieve $9\,\mathrm{s}$ of microgravity with a residual acceleration of $\sim 10^{-5}g_0$, where $g_0=9.81\,\mathrm{m~s^{-2}}$ is the Earth's gravitational acceleration. During the experiment the particle ensemble was captured by a high-speed, high-resolution CCD camera operated at 115 frames per second (fps) with a resolution of $1024\times 1024$ pixels (see Fig. \ref{f_dt_exp} for a sample image).

The design of the test cell allows for free collisions of the particles and the limitation of the height to $1.5$ particle diameters ensures an unambiguous analysis of their motion, because the samples cannot pass each other in the line of sight of the camera. In future experiments the studies will be extended to pure water ice samples in a cryogenic environment for a more realistic simulation of Saturn's dense rings.

\section{Results}\label{s_res}

In this work, we report on one out of seven experiments conducted with 92 solid glass beads ($1\,\mathrm{cm}$ diameter). Two sets of 32 glass spheres each were injected into the experiment chamber from opposites sides, while 28 particles were at rest inside the cell. After a short period of equilibration, the kinetic energy of the system decreased due to inelastic collisions. To analyze the particles' motion all images were convolved with an image of an individual sample sphere giving the most probable positions of all particles' center coordinates. Afterwards the images were binarized and a particle tracking algorithm was used to determine the trajectories of all particles. From this data, the individual sample velocities could be calculated for each image frame. This means that (although generally possible) the collisions were treated in a statistical way, rather than being analyzed individually.\par

\begin{figure}[!tb]
  \includegraphics[angle=90,width=\columnwidth]{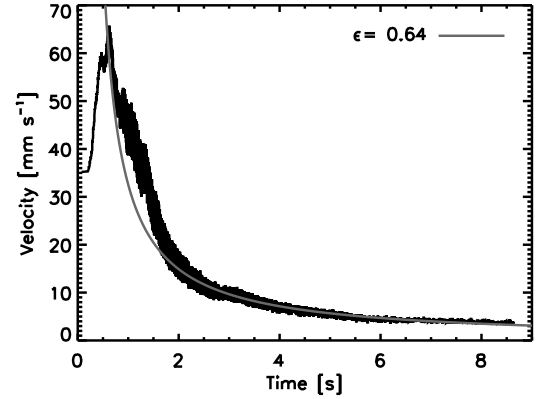}
  \caption{Velocity decay over experiment duration. After a short period of equilibration the velocity decay can be fitted by a simple equation which is in accordance with Haff's Law \citep{haff1983JFM}.}\label{f_dt_plot_haff}
\end{figure}

The results can be found in Fig. \ref{f_dt_plot_haff}, which shows that after the equilibration (first 2~s of experiment duration) the velocity decays systematically due to inelastic collisions and excitation of rotational motion. Assuming a constant coefficient of restitution $\varepsilon = const.$, describing the energy loss in individual collisions, the mean velocity $v$ as a function of time can be derived to be

\begin{eqnarray}
 v\left(t\right)&=&\frac{1}{\frac{1}{v_0}+(1-\varepsilon)\sigma n t}\; , \label{eq_vel}
\end{eqnarray}
\noindent
where $v_0$ is the initial injection velocity, $\sigma = 4\pi r^2$ is the collisional cross-section, and $n$ is the number density of the particle ensemble. A fit of Eq. (\ref{eq_vel}) to the data shows that the low-velocity regime (after approx. 2~s of experiment time) can be well described with a constant coefficient of restitution of $\varepsilon = 0.64$ (solid curve in Fig. \ref{f_dt_plot_haff}).\par

\begin{figure}[!tb]
  \includegraphics[angle=90,width=\columnwidth]{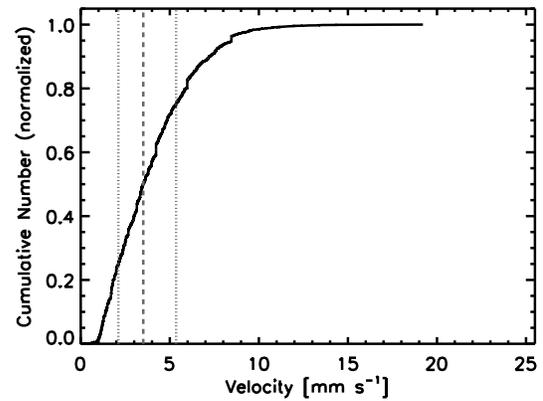}
  \caption{Normalized cumulative number of samples with velocity $\leq v$ for the time interval $6.5\ldots 8.5\,\mathrm{s}$. The mean particle velocity is $3.5\,\mathrm{mm\,s^{-1}}$ (dashed line) with 50\% of all samples covering the velocity range $2\ldots 5\,\mathrm{mm\,s^{-1}}$ (dotted lines). Hence, the collisions occur in the velocity regime relevant for studying Saturn's rings.}\label{f_dt_plot_cum}
\end{figure}

Additionally, we found that the mean velocity falls as low as $\sim 3.5\,\mathrm{mm\,s^{-1}}$ during the last 2~s of experiment duration with 50\% of all sample velocities being in the range $2\ldots 5\,\mathrm{mm\,s^{-1}}$ (see Fig. \ref{f_dt_plot_cum}).

\section{Summary}

We built an experimental setup that is suitable for studying free collision in a particle ensemble under microgravity conditions and we were able to perform prototype experiments with solid glass spheres. The particles could be tracked automatically and the development of the mean particle motion could be well fitted by Eq. (\ref{eq_vel}), yielding a constant coefficient of restitution of $\varepsilon = 0.64$. The achieved velocities are sufficiently low to simulate collisions within Saturn's dense main rings. As water ice samples are expected to dissipate more kinetic energy in collisions ($\overline{\varepsilon} = 0.45$; cf. \citep{heisselmann_et_al2009Icarus}) the velocities in future experiments are expected to be even lower.

\section{Future Work}

In future, similar experiments will be conducted using pure water-ice samples. Recent studies \citep{heisselmann_et_al2009Icarus} show that the mean coefficient of restitution of colliding cm-sized ice spheres is $\sim \overline{\varepsilon}= 0.45$. Hence, this gives rise to the question under which circumstances ice particles might coagulate (i.e. form aggregates).\par

Another key issue in understanding the stability and evolution of Saturn's rings will be the study of the role of fragmentation and the influence of regolith-covered surfaces in binary collisions. Additionally, the conditions (e.g. number density, velocity regime) for the formation of clumps, patterns and instabilities \citep{brilliantov_poeschel2004ktg, valverde_et_al2004PRL, spahn_et_al2000NoteIcarus} will be thoroughly investigated in our upcoming many-body experiments.\par

Planned diamagnetic-levitation studies of perturbations of particle conglomerates by shear forces will simulate Keplerian shear in Saturn's rings and are expected to shed light on the mechanisms leading to the formation of structures like the `propellers', predicted by theory \citep{spahn1987Icarus} and observed by the \emph{Cassini} spacecraft \citep{sremcevic_et_al2007Nature}.


\begin{theacknowledgments}

We are grateful to M. Böttger, S. Buschschlüter, S. Kothe, and M. Thede for their contribution to the development and conduction of the drop tower experiments. We also acknowledge R. Weidling for his efforts during the image processing.

This work was generously funded by the German Space Agency (DLR) under grant No. 50 WM 0636. We also thank DLR for providing the drop-tower flights for our studies.

\end{theacknowledgments}



\bibliographystyle{aipproc}   

\bibliography{../Literatur,../Unpublished}

\IfFileExists{\jobname.bbl}{}
 {\typeout{}
  \typeout{******************************************}
  \typeout{** Please run "bibtex \jobname" to optain}
  \typeout{** the bibliography and then re-run LaTeX}
  \typeout{** twice to fix the references!}
  \typeout{******************************************}
  \typeout{}
 }

\end{document}